\documentclass[12pt]{article}

\usepackage{scicite}

\usepackage{times}

\usepackage{amsmath}
\usepackage{amsfonts}
\usepackage{amssymb}
\usepackage{graphicx}

\newenvironment{sciabstract}{%
\begin{quote} \bf}
{\end{quote}}

\title{Division of labor, skill complementarity, and heterophily in socioeconomic networks}

\author
{Wen-Jie Xie,${}^{1,2,3}$ Ming-Xia Li,${}^{2,3}$ Zhi-Qiang Jiang,${}^{1,2}$ Qun-Zhao Tan,${}^{4}$ \\Boris Podobnik,${}^{5,6,7,8}$ Wei-Xing Zhou,${}^{1,2,3\ast}$ H. Eugene Stanley${}^{5\ast}$\\
\\
\normalsize{${}^{1}$Department of Finance, East China University of Science and Technology,}\\
\normalsize{Shanghai 200237, China}\\
\normalsize{${}^{2}$Research Center for Econophysics, East China University of Science and Technology,}\\
\normalsize{Shanghai 200237, China}\\
\normalsize{${}^{3}$Department of Mathematics, East China University of Science and Technology,}\\
\normalsize{Shanghai 200237, China}\\
\normalsize{${}^{4}$Shanda Games Ltd., 690 Bibo Road, Shanghai 201203, China}\\
\normalsize{${}^{5}$Department of Physics and Center for Polymer Studies, Boston University, MA, USA}\\
\normalsize{${}^{6}$Zagreb School of Economics and Management, 10000 Zagreb, Croatia}\\
\normalsize{${}^{7}$Faculty of Civil Engineering, University of Rijeka, 51000 Rijeka, Croatia}\\
\normalsize{${}^{8}$Faculty of Economics, University of Ljubljana, 1000 Ljubljana, Slovenia}\\
\\
\normalsize{$^\ast$To whom correspondence should be addressed; E-mail:  wxzhou@ecust.edu.cn, hes@bu.edu.}
}

\date{}

\begin{document}



\maketitle



\begin{sciabstract}
  Constituents of complex systems interact with each other and self-organize to form complex networks. Empirical results show that the link formation process of many real networks follows either the global principle of popularity or the local principle of similarity or a tradeoff between the two. In particular, it has been shown that in social networks individuals exhibit significant homophily when choosing their collaborators. We demonstrate, however, that in populations in which there is a division of labor, skill complementarity is an important factor in the formation of socioeconomic networks and an individual's choice of collaborators is strongly affected by heterophily. We analyze 124 evolving virtual worlds of a popular ``massively multiplayer online role-playing game'' (MMORPG) in which people belong to three different professions and are allowed to work and interact with each other in a somewhat realistic manner. We find evidence of heterophily in the formation of collaboration networks, where people prefer to forge social ties with people who have professions different from their own. We then construct an economic model to quantify the heterophily by assuming that individuals in socioeconomic systems choose collaborators that are of maximum utility. The results of model calibration confirm the presence of heterophily. Both empirical analysis and model calibration show that the heterophilous feature is persistent along the evolution of virtual worlds. We also find that the degree of complementarity in virtual societies is positively correlated with their economic output. Our work sheds new light on the scientific research utility of virtual worlds for studying human behaviors in complex socioeconomic systems.
\end{sciabstract}


\section*{Introduction}

Complexity emerges in the evolving and self-organizing processes of many natural, social, technological, and biological systems. The constituents of a complex system interact with each other and form complex evolving networks, where the constituents are nodes and their interaction relationships are links \cite{Albert-Barabasi-2002-RMP,Newman-2003-SIAMR,Boccaletti-Latora-Moreno-Chavez-Hwang-2006-PR,Fortunato-2010-PR,Estrada-Hatanoe-Benzi-2012-PR,Holme-Saramaki-2012-PR}. For many real networks, the link formation process follows either the global principle of popularity in which a node tends to link with high-degree nodes  \cite{Simon-1955-Bm,Barabasi-Albert-1999-Science}, or the local principle of similarity in which a node tends to link with nodes having traits similar to its own \cite{Papadopoulos-Kitsak-Serrano-Boguna-Krioukov-2012-Nature}, or a tradeoff between them \cite{Papadopoulos-Kitsak-Serrano-Boguna-Krioukov-2012-Nature}.

In the sociological literature the local principle of similarity, i.e., the phenomenon that ``birds of a feather flock together,'' is known as homophily \cite{McPherson-SmithLovin-Cook-2001-ARS}. There is much empirical evidence indicating that individuals prefer to forge social ties with people whose traits such as education, race, age, and sex are the same as their own \cite{Currarini-Jackson-Pin-2009-Em,Currarini-Jackson-Pin-2010-PNAS,Cheng-Xie-2013-PNAS,Kovanen-Kaski-Kertesz-Saramaki-2013-PNAS}. Such homophilous behaviors are ubiquitous in social networks and have been well documented \cite{Coleman-1958-HO,Moody-2001-AJS,McPherson-SmithLovin-Cook-2001-ARS,Kossinets-Watts-2009-AJS,Currarini-Jackson-Pin-2009-Em,Currarini-Jackson-Pin-2010-PNAS,Apicella-Marlowe-Fowler-Christakis-2012-Nature,Kovanen-Kaski-Kertesz-Saramaki-2013-PNAS}.  In addition, the similarity shared by individuals in a group is often a significant predictor of a group's altruism level and its ability to cooperate \cite{Curry-Dunbar-2013-HN}. Sociological literature argues that human societies tend to display two social systems: (i) homophilous, in which people seek out people who are similar, and (ii) heterophilous, in which people seek out people who are different \cite{Rogers-2003}. The evidence indicating the actual existence of heterophilous societies is rare, however.

In general, it has long been accepted that one of the most significant factors in increasing productivity in modern human societies has been the division of labor \cite{Smith-1776}. Thus we might assume that people in modern societies now prefer to forge links or collaborate with those who have complementary productive skills and that socioeconomic networks are becoming increasingly heterophilous, but no direct evidence of this has been documented. The availability of big data recorded from massively multiplayer online role-playing games (MMORPGs) enables us to test social and economic hypotheses and theories---such as this one---in large-scale virtual populations \cite{Bainbridge-2007-Science} and gain a deeper understanding of our social and economic behaviors \cite{Jiang-Zhou-Tan-2009-EPL,Szell-Lambiotte-Thurner-2010-PNAS,Szell-Thurner-2010-SN,Thurner-Szell-Sinatra-2012-PLoS1,Szell-Sinatra-Petri-Thurner-Latora-2012-SR,Szell-Thurner-2012-ACS,Klimek-Thurner-2013-NJP,Szell-Thurner-2013-SR,Fuchs-Sornette-Thurner-2014-SR}.

We study the evolving collaboration networks of 124 virtual worlds of a popular MMORPG in which the avatars (virtual people) belong to three different professions (warrior, mage and priest) with different skills. In a virtual world, each avatar can maintain a list of collaborators and the closeness of each pair of collaborators is quantitatively measured by their mutual intimacy. More details about the data are found in {\it{Material and Methods}} and {\textit{SI text}}. We find that avatars prefer to collaborate with others of different professions in these virtual worlds. We then propose an economic model to quantify the heterophily by assuming that individuals in socioeconomic systems choose collaborators that are of maximum utility. Model calibration shows that the evolving curves preference coefficient $\gamma_{ij}(t)$ that quantifies the degree of collaboration preference from $i$-avatars to $j$-avatars are well separated and illustrate remarkable heterophily: $\gamma_{ij}(t)>1$ for avatars belonging to two different professions ($i\neq{j}$) and $\gamma_{ij}(t)<1$ for avatars of the same profession ($i=j$). We further construct two measures to quantify the degree of complementarity of virtual societies. We find that social complementarity positively correlates with economic output.

\section*{Results}

\subsection*{Empirical analysis}

\begin{figure*}[tb]
\begin{center}
  \includegraphics[width=\linewidth]{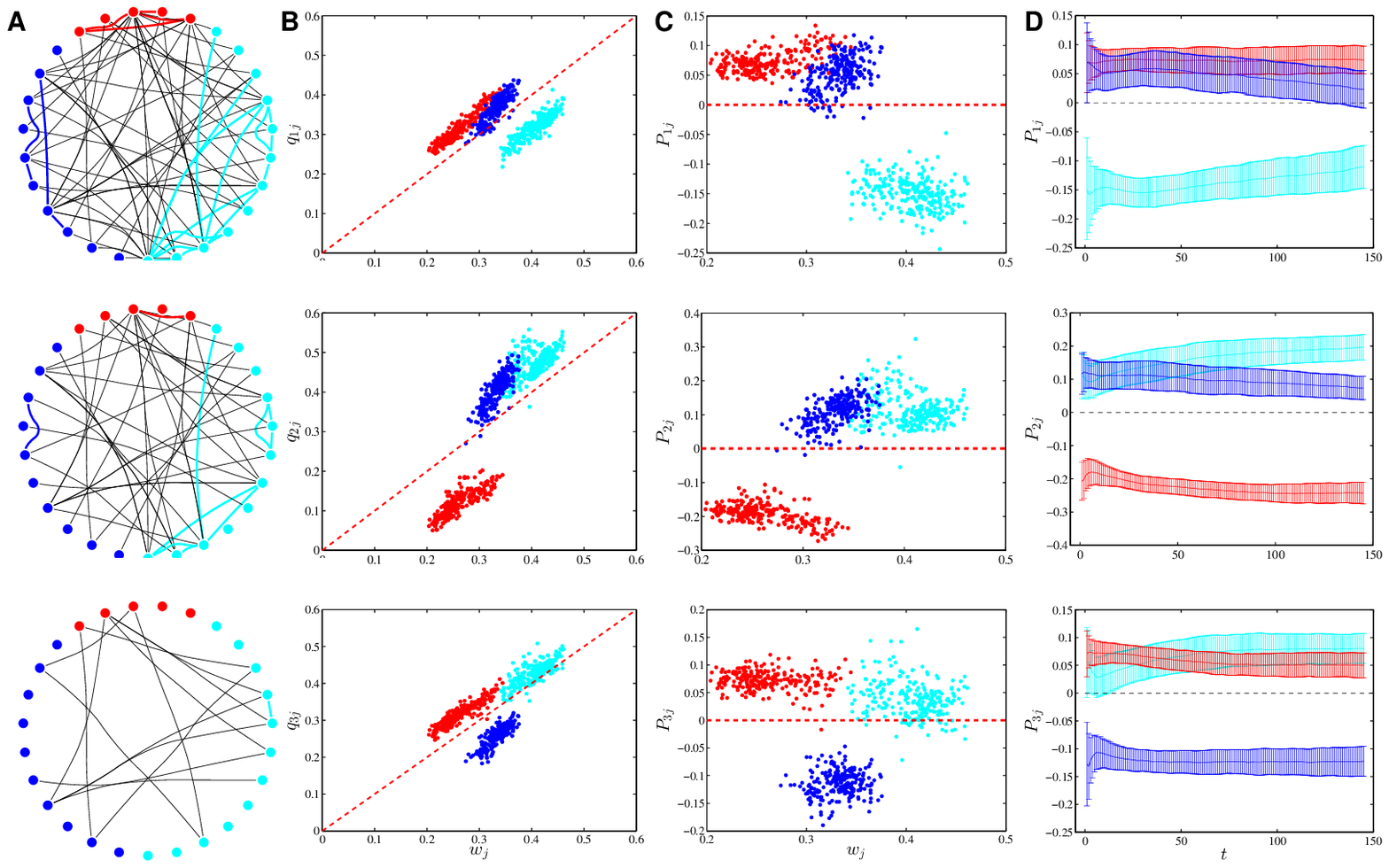}
  \caption{\label{Fig:PreferenceEmpirics} Empirical evidence of heterophily in the socioeconomic networks of virtual societies on a typical day $t=15$. Warriors, priests and mages are marked respectively in cyan, red and blue. ({\textit{A}}) Networks of 27 avatars randomly chosen from a virtual society filtered by three intimacy thresholds $I_c=0$, $100$ and $2000$ (top to bottom). ({\textit{B}}) Dependence of $q_{s,ij}$ on relative size $w_{s,j}$ for all virtual societies for $I_c=100$. In each plot, there are three well isolated clusters. For most societies, $q_{s,ij}> w_{s,j}$ when $i\neq j$ and $q_{s,ij}<w_{s,j}$ when $i=j$. ({\textit{C}}) Dependence of preference measure $P_{s,ij}$ on relative size $w_{s,j}$ for all societies for $I_c=100$. There are also three well separated clusters in each plot. For most societies, $P_{s,ij}>0$ when $i\neq j$ and $P_{s,ij}<0$ when $i=j$. ({\textit{D}}) Evolution of the averaged preference measure $P_{s,ij}$ over all virtual societies for $I_c=100$. The preference measures are roughly persistent.}
\end{center}
\end{figure*}

Fig.~\ref{Fig:PreferenceEmpirics}{\it{A}} shows the collaboration networks on day $t=15$ of a group of 27 avatars randomly chosen from a virtual society filtered by three intimacy thresholds $I_c=0$, 100, and 2000. There are 12 warriors, 5 priests, and 10 mages. If $i$-avatars are homophilous (neutral, heterophilous) in their collaboration-forging process, the proportion of links between $i$-avatars is greater than (equal to, less than) the square of the proportion of $i$-avatars (0.1975 for warriors, 0.0343 for priests, and 0.1372 for mages). For $I_c=0$, there are 77 links including 15 intra-warrior links, 4 intra-priest links, and 4 intra-mage links. The proportions of intra-profession links are 0.1948 for warriors, 0.0519 for priests, and 0.0519 for mages. For $I_c=100$, there are 48 links including 8 intra-warrior links, 1 intra-priest link, and 1 intra-mage link. The proportions of intra-profession links are 0.1667 for warriors, 0.0208 for priests, and 0.0208 for mages. For $I_c=2000$, there are 15 links including only one intra-warrior link and no intra-priest and intra-mage links. The proportions of intra-profession links are 0.0667 for warriors and 0 for priests and mages. Hence, the avatars in Fig.~\ref{Fig:PreferenceEmpirics}{\it{A}} are heterophious except for priests when $I_c=0$. We will show below that heterophily is not a specific characteristic for these 27 avatars but a universal feature present in all the virtual societies.


Consider a society $s$ on day $t$ whose size is the number $N_s$ of $s$-avatars. The number of $(s,i)$-avatars is denoted $N_{s,i}$, where $i=1$, 2, and 3 stand for the three professions. Hence $N_s=\sum_{i=1}^3N_{s,i}$. The ratio of $i$-avatars in society $s$ is
\begin{equation}
  w_{s,i}= {N_{s,i}}/{N_s}.
\end{equation}
The average number of $j$-collaborators of an $(s,i)$-avatar is $f_{s,ij}$. Hence, the average number of collaborators that an $(s,i)$-avatar has is $f_{s,i}=\sum_{j=1}^3 f_{s,ij}$. The average proportion of $j$-collaborators in all collaborators of an $i$-avatar is
\begin{equation}
   q_{s,ij}=\frac{f_{s,ij}}{\sum_{j=1}^3 f_{s,ij}}=\frac{f_{s,ij}}{f_{s,i}}.
\end{equation}
Note that $q_{s,ii}$ is the homophily index \cite{Coleman-1958-HO,Marsden-1988-SN}. If $i$-avatars have zero preference for collaborating with $j$-avatars, we have $q_{s,ij} = w_{s,j}$. If $i$-avatars prefer to collaborate with $j$-avatars, we have $q_{s,ij} > w_{s,j}$. In this case the $i$-avatars are homophilous when $j=i$ and the $i$-avatars are heterophilous when $j{\neq}i$. Fig.~\ref{Fig:PreferenceEmpirics}{\textit{B}} shows that when $t=15$ and $I_c=100$ most virtual societies have $q_{s,ij}>w_{s,j}$ when $i\neq j$, but $q_{s,ij}< w_{s,j}$ when $i= j$. Such heterophilous patterns are observed for other values of $t$ and $I_c$ as well.

Similar to the inbreeding homophily index \cite{Coleman-1958-HO,Marsden-1988-SN}, we define the collaboration preference index to be
\begin{equation}
   P_{s,ij}=\frac{q_{s,ij}-w_{s,j}}{1-w_{s,j}}.
\end{equation}
Note that $P_{s,ii}$ is the inbreeding homophily value \cite{Coleman-1958-HO,Marsden-1988-SN}. If $i$-avatars have no preference to collaborate with $j$-avatars, we have $P_{s,ij}=0$. If $i$-avatars prefer to collaborate with $j$-avatars, we have $P_{s,ij}>0$. In the latter case, the $i$-avatars are homophilous when $j=i$ and heterophilous when $j{\neq}i$. Empirical results show that for most virtual societies $P_{s,ij}>0$ when $i\neq j$, but $P_{s,ij}<0$ when $i= j$ (Fig.~\ref{Fig:PreferenceEmpirics}{\textit{C}}). Thus in socioeconomic networks the avatars are heterophilous.

Fig.~\ref{Fig:PreferenceEmpirics}{\textit{D}} shows the evolution of preference values $P_{ij}$ averaged over all societies on the same day for $I_c=100$. Although these curves exhibit mild trends, it is evident that the heterophilous feature is persistent as the virtual societies develop.

\subsection*{Quantifying collaboration preference}

Viewing virtual worlds as analogous to real societies, an avatar seeks collaborators when she finds it difficult to complete a task alone. The choice of collaborators has a significant influence on the completion of the task and it is better to have collaborators with complementary skills. Hence the number and skill configuration (or distribution) of an avatar's collaborators are the main determinants of her utility. If the skill configuration in the collaborator list of an avatar is optimal, her utility reaches its maximum. If the skill configuration deviates from that optimal value, her utility is reduced. Taking into account the fact that maintaining a social network incurs a cost, we propose an economic model for the decision-making process of an avatar to determine the optimal number of collaborators needed (see {\textit{Materials and Methods}}). The model essentially maximizes the function
\begin{equation}
 D_{s,i}(f_{s,i}) = (b-c)f_{s,i}^{\beta}-af_{s,i}^{\alpha} \left[\sum_{j=1}^3(q_{s,ij}-\gamma_{ij}w_{s,j})^2\right] ^{\frac{\alpha}{2}},
\end{equation}
where $\gamma_{ij}$ is the preference coefficient. If $i$-avatars prefer to collaborate with $j$-avatars, $\gamma_{ij}>1$; if $i$-avatars have no preference regarding $j$-avatars, $\gamma_{ij}=1$; if $i$-avatars prefer not to collaborate with $j$-avatars, $\gamma_{ij}<1$. For $\{i,j,k\}=\{1,2,3\}$, if $\gamma_{ij}>\gamma_{ik}$, then $i$-avatars prefer $j$-avatars over $k$-avatars.

\begin{figure}[thb]
\centering
  \includegraphics[width=0.50\linewidth]{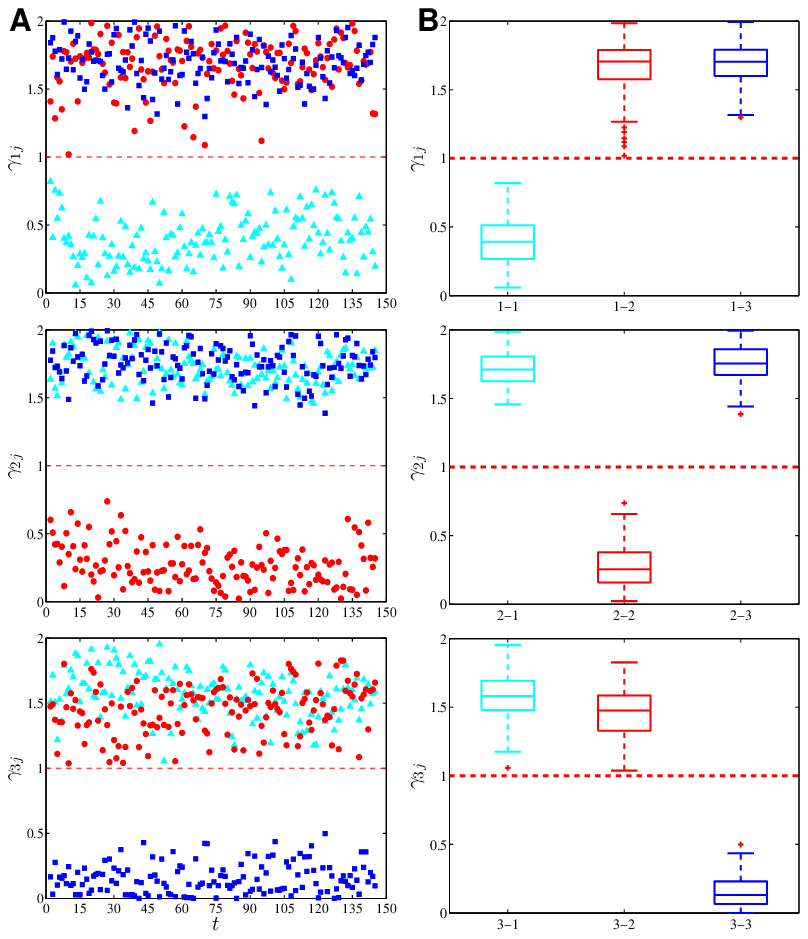}
  \caption{\label{Fig:PreferenceModel:Gamma} Preference coefficients $\gamma_{ij}$ for socioeconomic networks with the intimacy threshold being $I_c=100$. ({\textit{A}}) Daily evolution of the nine preference coefficients $\gamma_{ij}$ with $i,j\in\{1,2,3\}$. The color of a point $(t,\gamma_{ij})$ is determined by $j$: cyan, red and blue for $j=1$, 2 and 3, respectively. The nine points for a given $t$ were determined simultaneously in one calibration. ({\textit{B}}) Box plots of $\gamma_{ij}$ shown in {\textit{A}}. }
\end{figure}

To calibrate the model, we follow and further develop an econometric method presented in Ref.~\cite{Currarini-Jackson-Pin-2010-PNAS} (see {\textit{Materials and Methods}}). We obtain the values of $\gamma_{ij}$ for each intimacy threshold $I_c$ on each day $t$. Fig.~\ref{Fig:PreferenceModel:Gamma}{\textit{A}} shows the evolution of preference coefficients $\gamma_{ij}$ for socioeconomic networks using the intimacy threshold $I_c=100$, and Fig.~\ref{Fig:PreferenceModel:Gamma}{\textit{B}} shows the average preference coefficients over all days. The $F$-tests presented in {\textit{Materials and Methods}} show that all the results are significant at the 0.1\% level.

All the estimated values of the $\gamma_{ii}$ coefficients are less than 1, while all the $\gamma_{ij}$ values for $i\neq{j}$ are greater than 1. This indicates that the avatars are not seeking same-profession avatars but different-profession avatars and are thus heterophilous. In most cases, especially when the intimacy threshold $I_c$ is not large, the $\gamma_{ij}(I_c,t)$ values do not have a trend along the evolution of virtual worlds. When $I_c$ is large, however, we observe an increasing trend in $\gamma_{13}(I_c,t)$ for $I_c=1000$ and $2000$, in $\gamma_{23}(I_c,t)$ for $I_c=1000$ and $2000$, and in $\gamma_{32}(I_c,t)$ for $I_c=500$, $1000$ and $2000$. We find that the preference coefficients might change with the increase of $I_c$. For warriors, $\gamma_{11}$ and $\gamma_{13}$ decreases, while $\gamma_{12}$ increases. For priests, $\gamma_{21}$ increases, $\gamma_{22}$ does not exhibit evident trend, while $\gamma_{23}$ decreases. For warriors, $\gamma_{31}$ increases,  $\gamma_{32}$ decreases, while $\gamma_{12}$ increases for large $I_c$ values.

There are also intriguing patterns of relative collaboration preference as quantified by $\gamma_{ij}-\gamma_{ik}$ where $i$, $j$ and $k$ correspond to the three professions  (Fig.~\ref{Fig:PreferenceModel:Gamma}{\textit{B}}). On average, warriors prefer priests over mages and this relative preference enhances when $I_c$ becomes greater but reduces slightly when $t$ increases for large $I_c$ values. Priests prefer mages over warriors when $I_c$ values are small and prefer warriors over mages when $I_c$ values are large. For large $I_c$, priests' relative preference on warriors over mages decreases along time $t$. Mages prefer priests over warriors when $I_c$ is small and prefers warriors over priests when $I_c$ is large. For large $I_c$, mages' relative preference on warriors over priests also decreases along time $t$.


\subsection*{Group complementarity and economic output}

To measure the economic implications of heterophilous preference in socioeconomic networks, we investigate the relationship between complementarity of professions and economic performance. Consider the socioeconomic network ${\cal{N}}_s(I_c,t)$ of a virtual society with intimacy threshold $I_c$ on day $t$. Economic production utilizes virtual money and goods that are converted to a standardized currency (see {\textit{Materials and Methods}}). For each member avatar $a$ in ${\cal{N}}_s(I_c,t)$, we calculate her production output in the week from $t-6$ to $t$, denoted as $Y_{s,a}(t)$. The economic performance of the avatars in ${\cal{N}}_s(I_c,t)$ is defined as the output per capita,
\begin{equation}
   Y_s(I_c,t) = \frac{1}{\#{\cal{N}}_s(I_c,t)}\sum_{a\in{\cal{N}}_s(I_c,t)} Y_{s,a}.
\end{equation}
One measure of profession complementarity can be defined as the sum of preference measures between the three types of avatars,
\begin{equation}
  P_s(I_c,t)=\sum_{j\neq i}P_{s,ij}(I_c,t).
\end{equation}
Alternatively, we can measure complementarity by determining how much the real collaborator configuration $q_{s,ij}$ deviates from the optimal collaborator configuration $\gamma_{ij}w_{s,j}$ (see {\textit{Materials and Methods}}). The lower the deviation, the higher the degree of complementarity. Thus, we have
\begin{equation}
  C_s(I_c,t)\equiv\left[\sum_{i=1}^3 \sum_{j=1}^3
    (q_{s,ij}-\gamma_{ij}w_{s,j})^2\right]^{-\frac{1}{2}}.
\end{equation}

\begin{figure*}[tb]
\centering
  \includegraphics[width=\linewidth]{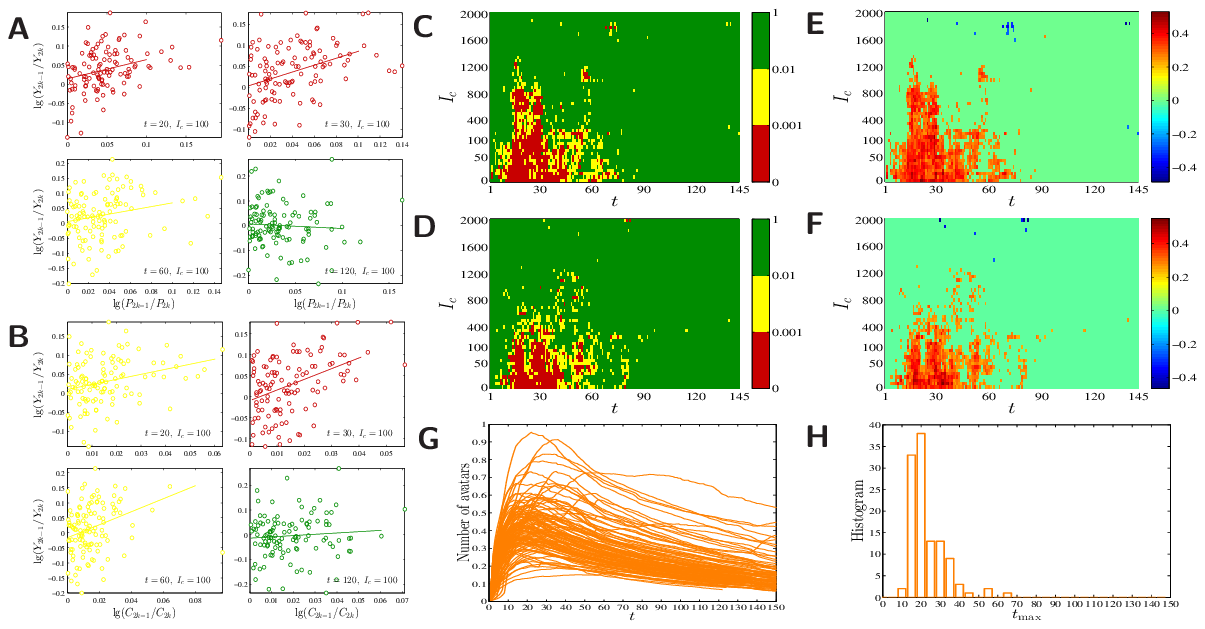}
  \caption{\label{Fig:Preference:Yield} Relation between complementarity of collaboration network and economic output. ({\textit{A}}) Examples of correlations between $\lg({P_{2k-1}}/{P_{2k}})$ and $\lg({Y_{2k-1}}/{Y_{2k}})$. ({\textit{B}}) Examples of correlations between $\lg({C_{2k-1}}/{C_{2k}})$ and $\lg({Y_{2k-1}}/{Y_{2k}})$. ({\textit{C}}) The $p$-value of the correlation between $\lg({P_{2k-1}}/{P_{2k}})$ and $\lg({Y_{2k-1}}/{Y_{2k}})$ for different values of $I_c$ and $t$. A give grid $(t,I_c)$ is colored as red or yellow if the correlation is significant at the 0.001 level or the 0.01 level. Otherwise, the grid is colored as green. ({\textit{D}}) The $p$-value of the correlation between $\lg({C_{2k-1}}/{C_{2k}})$ and $\lg({Y_{2k-1}}/{Y_{2k}})$. ({\textit{E}}) Correlation coefficient $\rho$ between $\lg({P_{2k-1}}/{P_{2k}})$ and $\lg({Y_{2k-1}}/{Y_{2k}})$ for different values of $I_c$ and $t$. The correlation coefficient is set to be zero is the correlation is insignificant at the 0.01 level. ({\textit{F}}) Correlation coefficient $\rho$ between $\lg({C_{2k-1}}/{C_{2k}})$ and $\lg({Y_{2k-1}}/{Y_{2k}})$. ({\textit{G}}) Evolution of the number of active avatars in different virtual worlds. ({\textit{H}}) Histogram of $t_{\max}$ which is the date that a virtual world has historically the maximum active avatars. }
\end{figure*}

To make these results comparable for different virtual worlds, we investigate the relative quantities between two societies in the same world, $\lg({P_{2k-1}}/{P_{2k}})$, $\lg({C_{2k-1}}/{C_{2k}})$ and $\lg({Y_{2k-1}}/{Y_{2k}})$, rather than focusing on each society separately. Both measures of complementarity correlate strongly with the relative economic output when $t$ and $I$ are not large (Fig.~\ref{Fig:Preference:Yield}{\textit{A}}-{\textit{F}}). For the first few days (small $t$), most avatars strive to achieve higher levels by implementing specific tasks with small economic outputs. Other avatars attempt to obtain high intimacy levels by killing monsters in locations unrelated to economic outputs. In both cases the avatars intend to form complementary collaboration networks, but their activities are not focused on economic outputs. With the development of a virtual world, the number of active avatars increases and reaches a maximum at time $t_{\max}$ and then decays (Fig.~\ref{Fig:Preference:Yield}{\textit{G}}). When the activity level of a virtual world decreases, the intent of the avatars moves away from production and the collaboration structure is increasingly unrelated to economic activities. This is consistent with the fact that the spectrum of $t_{\max}$ has a distribution similar to the significant correlations between complementarity and economic output (Fig.~\ref{Fig:Preference:Yield}{\textit{H}}).

\section*{Discussion}

Overwhelming empirical evidence has shown that most social networks are homophilous. The probability that two nodes will connect is higher if they share similar traits. Our analysis of virtual worlds in which division of labor is operative demonstrates the important role of complementarity. In those socioeconomic networks individuals have the motivation to cooperate, and in the formation of the network individuals exhibit a heterophilous preference for those with complementary productive skills. Although mapping human behavior in virtual worlds to real-world human behavior is a subtle process \cite{Williams-2010-CT}, we believe that they share an intrinsic commonality because avatars in virtual worlds are, in fact, controlled by real-world people. In particular, avatars consciously form teams to accomplish tasks more successfully and effectively. More generally, growing evidence shows significant similarities in the behaviors of online avatars and real-world humans \cite{Bailenson-Blascovich-Beall-Loomis-2003-PSPB,Castronova-2006-GC,Bainbridge-2007-Science,Yee-Bailenson-Urbanek-Chang-Merget-2007-CPB,Lofgren-Fefferman-2007-LID,Grabowski-Kruszewska-2007-IJMPC,Xie-Li-Jiang-Zhou-2014-SR,Bhattacharya-Dugar-2014-MS,Ball-Newman-2013-NS}

In reality, human's preference is multidimensional in their traits \cite{Cheng-Xie-2013-PNAS}. The situation in virtual societies is a little different. Indeed, the way people interact with each other has significantly changed from the old days, particularly due to the impact of internet. In modern time, people can meet through internet in the virtual world instead of physically getting together to dine, drink, and talk to forge ties. Personal traits become less important in virtual societies while avatars' profession skill is identified as a dominating trait in virtual society.

%
%
%

The economic model proposed in this work is different from the one in \cite{Currarini-Jackson-Pin-2010-PNAS}. The essential difference is in the assumption of the utility function. Also, we have used a modified method of model calibration. Moreover, our model allows us to determine not only if $i$-avatars are homophilous or heterophilous but also the preference of one type of avatars to any other type of avatars. Hence, our model is more general and can be applied to other systems.

The relationship between social networks and economic output has been studied previously. It has been found, for example, that the diversity of individual relationships within a community strongly correlates with the economic development of the community \cite{Eagle-Macy-Claxton-2010-Science} and is directly associated with higher productivity for both individuals and the community \cite{Bettencourt-Samaniego-Youn-2014-SR,Ortman-Cabaniss-Sturm-Bettencourt-2015-SciAdv}. Because, to date, detailed real data at the population level of societies have been unavailable, this correspondence between professional skill and economic performance has not been quantified. Here we have begun to fill this data gap and also to highlight the usefulness of virtual worlds in carrying out research in economics and sociology \cite{Bainbridge-2007-Science}. One direct indication of our findings is that if a team leader or a firm manager recruits new members according the complementarity of their skills, the team's productivity will increase and the firm's economic well-being grow.

\noindent \textbf{Supplementary Material} accompanies this paper at {\small {\tt http://www.scienceadvances.org/}}.

\section*{Materials and Methods}
%

\subsection*{Data description}
We use a huge database recorded from $K=124$ servers of a popular MMORPG in China to uncover the patterns characterizing virtual socioeconomic networks. In a virtual world residing in a server there are two opposing camps or societies. Two avatars can choose collaborators, and a measure of closeness called intimacy is assigned to the collaboration link. When two collaborators in the same society collaborate to accomplish a task, their intimacy level increases. Two avatars from different societies can also collaborate, but their intimacy level remains zero. Hence the social networks of the two camps are essentially separate. We can regard the two camps as two societies, thus giving us $S=248$ virtual societies. For convenience, $s=2k-1$ and $s=2k$ stand respectively for the two societies in the same virtual world $k$. Two avatars are defined as collaborators if they both are on the collaborator list and their intimacy exceeds $I_c$. We consider many temporal collaboration networks. On day $t$ in a virtual society $s$, a network ${\cal{N}}_s(I_c,t)$ is a network in which the intimacies of all edges are no less than a threshold $I_c$, which can be disconnected (Fig.~\ref{Fig:PreferenceEmpirics}{\emph{A}}).

In each society there are three professions (warrior, priest, and mage). We use subscripts 1, 2, and 3 to stand respectively for the three professions: warrior, priest and mage.  For simplicity, we define several notations as follows. An {\textbf{$s$-avatar}} is an avatar belonging to society $s$. An {\textbf{$i$-avatar}} is an avatar having profession $i$. Similarly, an {\textbf{$i$-collaborator}} is a collaborator having profession $i$. An {\textbf{$(s,i)$-avatar}} or  {\textbf{$(s,i)$-collaborator}} is an $i$-avatar or $i$-collaborator in society $s$.

\subsection*{An economic model of collaboration formation}
If there is no collaboration preference, the proportion of $(s,j)$-collaborators that an $(s,i)$-avatar has is identical to the proportion of $j$-avatars in the group, that is $q_{s,ij}=w_{s,j}$. Hence the number of $(s,j)$-collaborators of an $(s,i)$-avatar is $f_{s,ij}^{\rm{rnd}}=f_{s,i}w_{s,j}$. In a virtual society with a preset division of labor, an avatar prefers collaborators with complementary skills. We assume that, for an $(s,i)$-avatar, there is an optimal configuration of collaborators with different skills, $f^{\rm{opt}}_{s,ij}=\gamma_{ij}f_{s,i}w_{s,j}$, where the preference coefficients $\gamma_{ij}$ are independent of society $s$. The utility of an $(s,i)$-avatar increases when her/his real number $f_{s,ij}$ of $(s,j)$-collaborators approaches the optimal value $f^{\rm{opt}}_{s,ij}$ and reaches its maximum $U_{s,i}^{\max}=bf_{s,i}^{\beta}$ when her/his collaborator configuration is optimal such that $f_{s,ij}=f^{\rm{opt}}_{s,ij}$. According to the law of diminishing marginal utility, we have $\beta<1$. Therefore, the utility function of an $(s,i)$-avatar is
\begin{equation}
   U_{s,i}
   =b_sf_{s,i}^{\beta}-a_sf_{s,i}^{\alpha} X_{s,i}^{\beta-\alpha},
\end{equation}
where
\begin{equation}
   X_{s,i} =\left[\sum_{j=1}^3(q_{s,ij}
     -\gamma_{ij}w_{s,j})^2\right]^{\frac{\alpha}{2(\beta-\alpha)}}
\end{equation}
and $\gamma_{ij}$ is the preference of $(s,i)$-avatars for $(s,j)$-avatars. If $i$-avatars do not have any preference on $j$-avatars such that $q_{s,ij}=w_{s,j}$ for all societies, we have $\gamma_{ij}=1$. If $i$-avatars prefer $j$-avatars, we have $\gamma_{ij}>1$. If $i$-avatars prefer not to collaborate with $j$-avatars, we have $\gamma_{ij}<1$. For $\{i,j,k\}=\{1,2,3\}$, if $\gamma_{ij}>\gamma_{ik}$, then $i$-avatars prefer $j$-avatars over $k$-avatars. To maintain a collaboration network of size $f_{s,i}$, the $(s,i)$-avatar suffers a cost proportional to $f_{s,i}$ \cite{Currarini-Jackson-Pin-2010-PNAS},
\begin{equation}
   M_{s,i}=c_sf_{s,i}^{\beta}.
\end{equation}
According to the above model, the overall utility in the decision-making process is
\begin{equation}
 D_{s,i}(f_{s,i})=U_{s,i}-M_{s,i}=(b_s-c_s)f_{s,i}^{\beta}-a_sf_{s,i}^{\alpha}
 X_{s,i}^{\beta-\alpha}.
\end{equation}

\subsection*{Model calibration}
An $(s,i)$-avatar solves the following decision-making problem of how many collaborators to have
\begin{equation}
  \max_{f_{s,i}} D_{s,i}(f_{s,i}).
\end{equation}
It follows that
\begin{equation}
   f_{s,i}=\left[{a_s\alpha}/{(b_s-c_s)\beta}\right]^{\frac{1}{\beta-\alpha}}
   X_{s,i}.
   \label{Eq:Solution:f:g:i}
\end{equation}
Note that the $\gamma_{ij}$ values are affected only by the professions and remain the same for different societies. This enables us to estimate the parameters.

The solution (\ref{Eq:Solution:f:g:i}) denotes the average behavior (decision) of all avatars having the same profession in a given society. If we consider an arbitrary avatar $a$, we must add a noise term \cite{Currarini-Jackson-Pin-2010-PNAS},
\begin{equation}
   f_{s,i;a}=\left[{a_s\alpha}/{(b_s-c_s)\beta}\right]
   ^{\frac{1}{\beta-\alpha}}X_{s,i} + \varepsilon_a,
   \label{Eq:Solution:f:g:i:a}
\end{equation}
which means that the ``realized'' number of collaborators avatar $a$ has is the sum of a universal (or systemic) term and an idiosyncratic error term. The error term is assumed to have mean 0 and variance $\sigma^2$. Note that this assumption states that the variance of any avatar of any profession is the same.

We denote $N_s$ as the size of society $s$ and $w_{s,i}$ as the fraction of $i$-avatars in society $s$. Hence the number of $i$-avatars in society $s$ is $N_s w_{s,i}$, and the expectation of the aggregated number of collaborators that $i$-avatars have in society $s$ is $N_s w_{s,i}f_{s,i}$. According to Eq.~(\ref{Eq:Solution:f:g:i:a}), we have
\begin{equation}
   N_s w_{s,i}f_{s,i}= N_s w_{s,i}
   \left[\frac{a_s\alpha}{(b_s-c_s)\beta}\right]^\frac{1}{\beta-\alpha}
   X_{s,i} +E_{s,i},
\label{Eq:wikNkTaik}
\end{equation}
where $E_{s,i}$ has mean 0 and variance $\left(N_sw_{s,i}\sigma\right)^2=\left(N_s w_{s,i}\right)^2\sigma^2$.

It follows that, for $i\neq{j}$,
\begin{equation}
  \frac{w_{s,i}N_sf_{s,i}-E_{s,i}}{w_{s,i}N_s X_{s,i}}=
     \frac{w_{s,j}N_sf_{s,j}-E_{s,j}}{w_{s,j}N_s X_{s,j}}.
\end{equation}
For society $s$, the error is
\begin{eqnarray}
  \Psi_{s,ij}&=&w_{s,i}N_sf_{s,i}w_{s,j}N_sX_{s,j} -w_{s,j}N_sf_{s,j}w_{s,i}N_sX_{s,i}\nonumber\\
             &=&N_s^2w_{s,i}w_{s,j}(f_{s,i}X_{s,j} -f_{s,j}X_{s,i})\\
             &=&E_{s,i}w_{s,j}N_sX_{s,j}-E_{s,j}w_{s,i}N_sX_{s,i}.
\label{Eq:Phi:E}
\end{eqnarray}
According to Eq.~(\ref{Eq:Phi:E}), we find that the mean of $\Psi_{s,i,j}$ is 0 and the variance is $\phi_{s,ij}\sigma^2$, where
\begin{eqnarray}
   \phi_{s,ij}
   &=& (w_{s,i}N_s)^2\left(w_{s,j}N_sX_{s,j}\right)^2 +(w_{s,j}N_s)^2\left(w_{s,i}N_sX_{s,i}\right)^2 \nonumber\\
   &=& N_s^4w_{s,i}^2w_{s,j}^2\left[\left(X_{s,i}\right)^2 + \left(X_{s,j}\right)^2\right]
\end{eqnarray}
Thus the normalized variable $\Psi_{s,ij}^2/\phi_{s,ij}$ has mean 0 and variance $\sigma^2$ for any society $s$. The sum of squared errors ($Q_{i,j}^2$) over all societies in the sample is
\begin{equation}
  Q_{ij}^2=\sum_{s=1}^S\frac{\Psi_{s,ij}^2}{\phi_{s,ij}}=\sum_{s=1}^S\frac{\left(f_{s,i}
    X_{s,j} -f_{s,j} X_{s,i} \right)^2} { \left(X_{s,i}\right)^2 +
    \left(X_{s,j}\right)^2 },
\end{equation}
which is independent of $N_s$ as expected. However, $Q_{ij}^2$ is dependent on $w_{s,i}$, which is consistent with the setup of our model but different from the model in
Ref.~\cite{Currarini-Jackson-Pin-2010-PNAS}. Thus the total sum of the squared errors is
\begin{equation}
  Q^2=Q_{12}^2+Q_{13}^2+Q_{23}^2.
\end{equation}
One can see that $a_s$, $b_s$ and $c_s$ could be society-specific and are not included in the final objective function of model calibration.

For each pair of $I_c$ and $t$, a society is excluded in model calibration if the number of avatars having at least one collaborator is less than 500 to ensure that $\epsilon_a$ has enough realizations. Changing this threshold around 500 results in same results. In addition, if the number of societies included in a model is less than 50, we do not calibrate the model because the model has 10 parameters.


To find the solution to the minimization of $Q^2$, the taboo search algorithm is adopted \cite{Cvijovic-Klinowski-1995-Science}. The solution space is restricted to $0 \leq\gamma_{ij}\leq2$ for $i,j\in \{1,2,3\}$ and $\frac{\alpha}{2(\beta-\alpha)}>0$. Because there are 10 free parameters, it is not easy to reach the global minimum. We thus perform a taboo search in each cell of a 9-dimensional lattice of size $2^9$ with the constraint that $0 \leq\gamma_{ij}\leq1$ or $1\leq\gamma_{ij}\leq2$. The parameters in certain cell corresponding to the minimum of $Q^2$ in all cells are obtained as the solution. The normality assumption of fitting errors has been verified by QQ-plots, which rationalizes the setup of the model. We note that the partitioning of the solution space into a 9-dimensional lattice of size $2^9$ is very important. If we perform the taboo search directly, the resulting $Q^2$ value is significantly larger and the three preference curves $\gamma_{ij}(t)$ for each $i$ are not well separated around $\gamma_{ij}=1$ (cf. Fig.~\ref{Fig:PreferenceModel:Gamma}).

\subsection*{Significance tests}
To test whether the preference coefficient $\gamma_{ij}$ of $i$-avatars to $j$-avatars is significantly different from the no-preference case, we perform $F$-tests using the null hypothesis
\begin{equation}
    H_0:~~\gamma_{ij}=1,~~~~~i,j\in\left\{1,2,3\right\}.
\end{equation}
Following Ref.~\cite{Currarini-Jackson-Pin-2010-PNAS}, the $F$-statistic
is
\begin{equation}
    F =
    {\frac{SSR_{\textrm{con}}-SSR_{\textrm{uncon}}}{p_{\textrm{uncon}}
        -p_{\textrm{con}}}}\bigg/{\frac{SSR_{\textrm{uncon}}}{n-p_{\textrm{uncon}}}},
\end{equation}
where $SSR$ is the sum of squared residuals of the best-fit calibration, $p$ is the number of model parameters, $n$ is the number of observations, while the subscript ``con'' indicates the constrained model under the null hypothesis and the subscript ``uncon'' the unconstrained model.

\subsection*{Economic output of individuals}
There are two virtual currencies, {\emph{Xingbi}} and {\emph{Jinbi}}. {\emph{Xingbi}} cannot be produced by an avatar's activity and can only be bought from the system, which has an approximately stable exchange rate in reference to the Chinese currency {\emph{Renminbi}}. {\emph{Xingbi}} is thus a universal currency across different virtual worlds.  {\emph{Jinbi}}, on the other hand, is produced by the economic activities of the avatars. There is a built-in exchange platform in each virtual world so that avatars can exchange {\emph{Xingbi}} and {\emph{Jinbi}}. In this way, there is a real-time exchange rate from {\emph{Jinbi}} to {\emph{Xingbi}}.

An avatar can produce virtual items (e.g., weapons, clothes, and medicines) and a limited amount of the virtual currency {\emph{Jinbi}}. We convert the produced items and {\emph{Jinbi}} to {\emph{Xingbi}} to obtain the real economic output of each avatar on each day. There is a marketplace in each virtual world in which avatars can sell their items that are priced in {\emph{Xingbi}} or {\emph{Jinbi}}. The price of an item is determined by the average price of all the trades in the marketplace on a given day. Each produced item can thus be measured in {\emph{Xingbi}}.



\noindent \textbf{Acknowledgements:}
%
The authors thank Didier Sornette for helpful conversations.


%
%

\end{document}